# Tailoring the Nucleation and Growth of Silver Nanoparticles by Sputtering Deposition under Acoustic Wave Activation. Assessment of Plasma Conditions and 2D Patterning Phenomena

Helene Reichel [a], Aurelio García-Valenzuela [a,†], José Andrés Espino-Román [a], Jorge Gil-Rostra [a], Guillermo Fernando Regodón [a], Víctor Rico-Gavira [a], Ana Borrás [a], Ana Gómez-Ramírez [a,b], Alberto Palmero [a], Agustín R. González-Elipe [a*], Manuel Oliva-Ramírez [a,b*].

[a] Laboratory of Nanotechnology on Surfaces and Plasma. Instituto de Ciencia de Materiales de Sevilla (CSIC-Universidad de Sevilla), Avda. Américo Vespucio 49, E-41092 Seville, Spain.

[b] Departamento de Física Atómica, Molecular y Nuclear, Universidad de Sevilla, Avda. Reina Mercedes, E-41012 Seville, Spain

[†] Current address: Ion Beam Physics and Materials Research, Helmholtz-Zentrum Dresden-Rossendorf, Bautzner Landstrasse 400, 01328 Dresden, Germany.

*corresponding authors: arge@icmse.csic.es; moliva1@us.es

**Abstract**

Early results on the plasma deposition of dielectric thin films on acoustic wave (AW) activated substrates revealed a densification pattern arisen from the focusing of plasma ions and their impact on specific areas of the piezoelectric substrate. Herein, we extend this methodology to tailor the plasma deposition of metals onto AW-activated $LiNbO_3$ piezoelectric substrates. Our investigation reveals the tracking of the initial stages of nanoparticle (NP) formation and growth during the submonolayer deposition of silver. We elucidate the specific role of AW activation in reducing particle size, enhancing particle circularity, and retarding NP agglomeration and account for the physical phenomena making these processes differ from those occurring on non-activated substrates. We provide a comparative analysis of the results obtained under two representative plasma conditions: diode DC sputtering and magnetron sputtering. In the latter case, the AW activation gives rise to a 2D pattern of domains with different amounts of silver and a distinct size and circularity for the silver NPs. This difference was attributed to the specific characteristics of the plasma sheath formed onto the substrate in each case. The possibilities of tuning the plasmon resonance absorption of silver NPs by AW activation of the sputtering deposition process are discussed.

**Introduction**

Acoustic waves (AWs) generated by the electroacoustic excitation of piezoelectric materials are currently used for many applications in a large variety of fields. Handling of liquids in microfluidics,[1,2] controlled separation of cells and nanoparticles from suspensions,[3–6] materials and metamaterials actuation,[7,8] activation of de-icing processes,[9–11] and many others[12] are examples of the rich variety of technical possibilities offered by this technology.

Among these applications, AWs and surface AWs (SAWs) have been used for the synthesis of materials and nanomaterials in liquid phase, with examples referring to the synthesis of silver nanoparticles (NPs) and nanostructures in microfluidic devices,[13] their mediation by the addition of DNA [14] or the promotion of their self-assembly in more complex structures.[3] However, the use of AW to activate the synthesis of nanostructures or thin films in vacuum or plasma environments has only recently been explored. An early but not sustained work by Takahashi et al.[15] reported the SAW-activated growth of Ni thin films to modify their magnetic properties. More recently, it has been realized that ultrasounds may activate a silica support during the silver ion implantation and lead to a modification of the size and form of the embedded silver nanoclusters.[16] Furthermore, we have also shown that the interaction of AWs with plasmas during the magnetron sputtering (MS) deposition of dielectric thin films drastically affects the deposition processes.[17,18] These effects have been explained by considering the appearance of a polarization pattern on the piezoelectric surface linked to the AW and its interaction with the plasma, resulting in the generation of a non-homogeneous plasma sheath and the focusing of impinging plasma ions on certain spatial regions of the substrate.[19] Patterning effects have been recently reported for plasma enhanced chemical vapor deposition by atmospheric pressure plasmas generated in a dielectric barrier discharge reactor, where local modifications in the distribution of plasma filaments appears to be the main cause of the heterogenous deposition. [20] The initial nucleation of ad-atoms deposited by physical vapor deposition (PVD) methods, either evaporation or sputtering, the subsequent formation of nanoparticles (NPs) and their eventual coalescence into large islands is of paramount importance for the control of thin film properties and, as such, has deserved many sound experimental and theoretical studies over the years.[21–24] In this context, formation, growth, and coalescence of silver nuclei have been recurrently studied under the influence of a large variety of physical variables and activation interactions, including temperature, density, the type and distribution of surface defects, and other energy activation processes. From a fundamental point of view, research has focused on analyzing or simulating basic physical phenomena, such as ad-atom diffusion, agglomeration, Ostwald ripening, and other physical processes taking place when metal ad-atoms are deposited on weakly interacting substrates.[25–27] The application works derived from these fundamental studies have pursued the determination of the best deposition conditions to achieve a strict control over particle size and shape and, in this way, gain an effective control over final properties of silver NPs and films such as optical and plasmonic behavior [28–31] or antimicrobial activity, two areas where the use of silver has raised most interest during the last years.[32–34]

The aim of this work is to explore the effects of AW activation during the first stages of nucleation and growth of metal NPs deposited by sputtering in the presence of plasma. Our previous work on dielectric thin films (i.e., $TiO_2$ and $SiO_2$) grown on AW activated substrates revealed a 2D patterning of their density and composition when they were plasma deposited by magnetron sputtering (MS) and demonstrated that this patterning replicated the polarization domains associated to the standing AW used for excitation.[17,18] In a leap forward, the present investigation addresses the sputtering deposition of metallic silver in the presence of plasmas under the AW activation of a $LiNbO_3$ piezoelectric substrate to study the initial stages of nucleation and NP growth. Experiments have been carried out using diode sputtering (DS)[35] or magnetron sputtering (MS)[36,37] to determine whether different plasma conditions may affect the deposition process. In both cases, the silver NPs were smaller and had a less tendency to agglomerate than those without AW activation. Remarkably, the appearance of a 2D patterned distribution of NPs was restricted to MS deposition, a feature that provides experimental clues about the plasma-AW interaction conditions required to give rise to this patterning phenomenon.

This investigation has contributed to shed light into the understanding of the physical processes involved upon AW activation of substrates. In addition, it has been possible to establish a series of principles to control the nucleation and growth processes of silver NPs prepared by sputtering. The disclosed mechanisms offer new engineering possibilities to adjust the size, shape, and 2D distribution of silver NPs, establishing a series of principles that can be straightforwardly transferred to practical applications where it is required a strict control over particle morphology and properties.

**Experimental**

*Silver NP and thin film deposition*

Silver has been deposited by plasma sputtering using two different experimental methods to disclose how NP growth is influenced by the interaction between the AW field and the plasma and to determine possible modifications due to the plasma characteristics. Selected plasma sputtering methods were the diode sputtering (DS) and the more widely utilized magnetron sputtering (MS). These two methods operate at two different pressures and therefore entail a different mean free path of particles within the plasma gas. They also differ in the thickness of the plasma sheath generated on the substrate surface. [35]

The DS deposition device consisted of a silver filament (15 cm long) negatively polarized at a voltage of 400 V in the interior of a cylindrical chamber where the sample holder was introduced with the substrates facing the filament at a distance of 2 cm. A scheme and photographs of this system are reported as supporting information, **Figure S1**. In this device, a stable Ar plasma was maintained at a pressure of $1.7 \cdot 10^{-1}$ mbar. The sputtering yield in this system was rather low and long deposition times in the order of tenths of minutes were required to reach deposited amounts of silver similar to those obtained after 1-2 min by the second procedure (i.e. magnetron sputtering (MS)).

The MS procedure consisted of a DC-pulse MS deposition system, making use of an oblique angle deposition (OAD) geometry. In this configuration, the substrate is rotated with respect to the target surface (with substrate rotation of $85^0$ in our case). Unlike the common use of this geometrical configuration to get nanocolumnar thin films,[38,39] in this work this oblique configuration did not aim to control a thin film nanostructure, but to reduce the arrival rate of silver atoms onto the surface and, in this way, to have a better control over submonolayer coverages. A circular 2" magnetron head working under the following conditions was used for the experiments: an Ar pressure of $5\cdot 10^{-3}$ mbar, a DC pulse frequency of 120 kHz, and an applied power of 25 W. Experiments were carried out for deposition times ranging from 30 s to 8 min. For comparative purposes and to have a better control over the first stages of NP growth, only samples deposited for 1 and 2 min were selected for analysis because they depicted an amount of deposited silver and a surface coverage similar to those reached upon deposition during tenths of minutes using the DS technique.

### *Deposition of silver NPs under AW activation*

Deposition under AW activation was carried out on transparent LiNbO$_3$ single crystal plates (20x10x0.5 mm) cut through the $01\bar{1}\bar{2}$ crystallographic plane (128$^0$ Y-cut). These plates were provided by Roditi International Corporation Ltd. The plates were held in a specially designed sample holder, where damping and possible attenuation of the AWs were minimized. Essentially, the LiNbO$_3$ plate was back-contacted with two metal electrodes (1x0.5 cm) placed on the two sides of the plate, according to a lateral field excitation (LFE) configuration. [40–42] This way, the upper surface of the plate was free and could be directly exposed to the sputtering source. It has been demonstrated that electroacoustic activation using this configuration generates standing AWs that propagate perpendicularly to the electrodes all along the piezoelectric substrate. These AWs have a predominant shear character (i.e. corresponding to a thickness shear mode (TSM) [43]) with a wavelength in the order of 1mm.[9] A full description of this holding piece and metal electrode can be found in previous works.[9,17,18] Moreover, and for comparative purposes, a reference LiNbO$_3$ substrate not subjected to AW activation and/or silicon or quartz substrates were also placed on the holder.

The two metallic contacts on the back side of the plate were used for electroacoustic excitation through the application of a sinusoidal voltage AC signal (112 V and 10 V amplitude for the DS and MS experiments, respectively) with a frequency in the range of 3.4 MHz, matching the resonance conditions of the plate. A schematic description of the holder setup, its location, deposition geometry, substrate and electrodes, and other experimental details can be found in the supporting information, **Figure S2** where a scheme of the AW generation and displacement current is also included. Further details about this experimental setup and the electronic system utilized for activation have been described in previous works.[9,17,18]

To ensure that proper AW resonance conditions are selected for the experiments, return loss spectra (i.e. $S_{11}$ parameter) were recorded with a vector network analyzer (VNA, SDR-Kits, model VNA 3SE) prior and during silver deposition. Since AW activation for long periods of time induced certain drifts in the resonance frequency likely because of an increase in temperature,[44] readjustments of that frequency were occasionally required. An example of such a drift is reported in the supporting information, **Figure S3**. Due to the long deposition times required for the experiments, this effect could be significant by DS deposition, but it was almost or negligible during the short periods of silver deposition by MS.

Along the text, the samples selected for characterization and analysis will be designated as Dt, D$_{ref}$t, MSt and MS$_{ref}$t, where *ref* means reference (i.e. without AW activation), D and MS diode and magnetron sputtering correspondently, and *t* stands for the deposition time utilized in each case.

*Characterization of silver NPs*

The deposited silver particles were optically characterized by UV-vis transmission spectroscopy in normal configuration in a Cary 100 instrument. In the MS samples an optical microscope (Olympus SZ40 Stereomicroscope) was also utilized to characterize the unbalanced distribution of silver NPs in the different submillimeter domains developed in this case.

Particle size, shape and distribution were characterized by Scanning electron microscopy in a Hitachi S4800 microscope provided with a field emission source (typically working with an acceleration potential of 2 kV) and an EDX detector capable of mapping the silver distribution over the surface (EDX Bruker-X Flash-4010). Both optical and SEM images were analyzed by the *ImageJ* software.[45] The SEM images were converted into black and white maps by applying a grey-scale threshold to determine the contours of the particles. The morphological operator *Opening* was applied to the images to identify adjacent distinct particles. The circularity of the particles was determined using the formula *circularity = 4π(area/perimeter^2),* and the particle sizes were estimated from their areas by approximating the shape of the particles to circles and determining their diameters.

**Results**

***AW-activated silver deposition and film formation***

The deposition of silver onto the AW-activated substrates was monitored following the evolution of the return loss spectra recorded with the VNA during the deposition. This evolution was registered for the MS experiments, where the deposition rate is high enough to ensure a proper control of the deposition process within a reasonably short period of time. **Figure 1a)** shows a series of return spectra recorded while depositing silver onto the $LiNbO_3$ substrate. To record the $S_{11}$ spectra, the VNA applies a very small AC voltage (200 mV) that is not high enough to induce any pattering or change in the growing mode of NPs. Curves with similar shapes have been reported for the deposition of silver on a quartz substrate activated with AW via a trice polar Antenna Transducer Acoustic Resonance (ATAR) setup AC excited with a signal of ca. 1 MHz frequency.[46] The thickness excitation (TE) of quartz crystal microbalance is known to proceed through TSM waves similar to those formed during LFE of piezoelectric substrates.[43] A shift in frequency towards lower values and a broadening and decrease in intensity until the complete neglect of the resonant $S_{11}$ peak are the main changes observed as a function of the deposition time as visualized in **Figure 1b**). In this figure, frequency and intensity are plotted as a function of the deposition time. We link the shape in the $S_{11}$ plot in **Figure 1b**) to successive silver deposition steps. First, an initial deposition step characterized by a constant resonance frequency and intensity of the $S_{11}$ band (plateau in Figure 1b)) where silver nuclei have not yet coalesced as illustrated in panel i) of Figure 1 c). When particle coalescence starts, roughly after 8 min of deposition, $S_{11}$ peak progressively shifts and broadens (i.e., in Figure 1b), and c) and panels ii) and iii)). For the experimental conditions in this work, the break point between isolated silver NPs and a continuous film occurs for a thickness of silver aggregates of approximately 20 nm (see **supporting information S4**). At this point, the formation of a continuous highly conductive film of percolated silver particles makes the surface of the substrate equipotential and prevents its polarization, leading to the complete banishment of the piezoelectric response of the activated substrate and the loss of their piezoelectric properties (i.e., the $S_{11}$ minimum disappears). We can therefore conclude that this type of $S_{11}$ curves can be used to monitor the progressive coverage of the piezoelectric substrate by silver particles and aggregates and their coalescence up to the build-up of a continuous metal film.

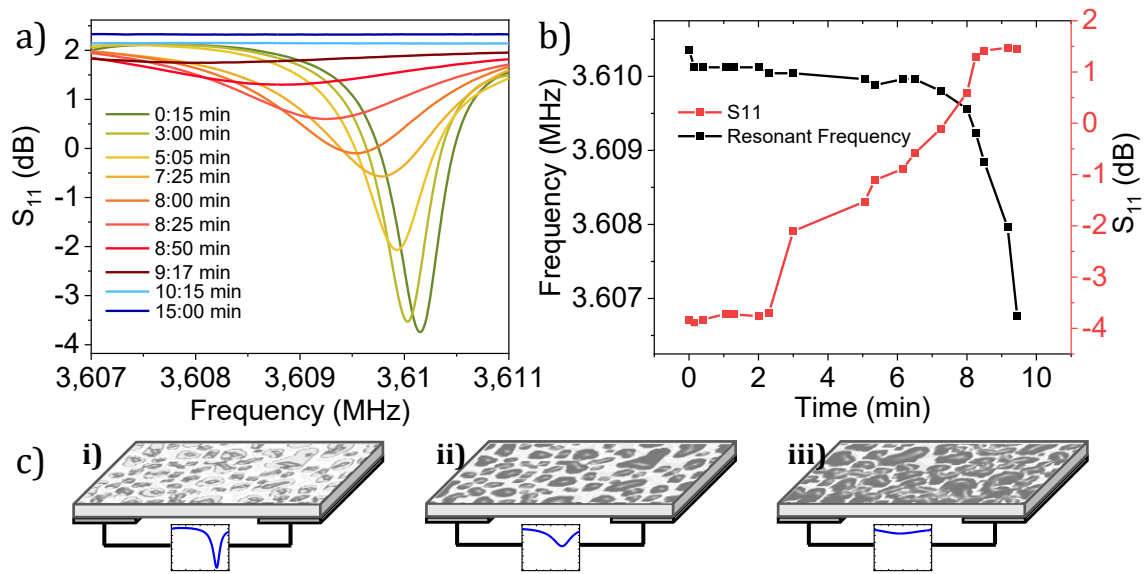

**Figure 1**.- *Evolution of the piezoelectric resonance for increasing Ag deposition times by MS.* a*) Plot of the $S_{11}$ spectra for different deposition times. b) Evolution of the $S_{11}$ peak frequency and intensity for increasing deposition times. c) Pictorial representation of the growth of the Ag nanoparticles of the LiNbO$_3$ plates depicting: i) early particle deposition stages (0-6 min); ii) particle growth stage (6-8 min); coalescence of the particles (8-9.3 min).*

For the purposes of the present work, this experiment provides a direct assessment of the amount of deposited silver and the evolution of its coalescence state. In particular, the practically constant values of intensity and frequency of $S_{11}$ up to 2 min of deposition provides a time window where silver should stay in the form of isolated NPs and the piezoelectric properties of the LiNbO$_3$ plate are not affected. Taking this into account, we have selected 1 and 2 min of deposition time for the MS experiments. In the DS experiments, similar but less defined changes in the $S_{11}$ parameter due to thermal drifting were found for longer deposition times between 16 and 40 min.

***Silver NP formation by DS deposition under AW activation.***

First, silver deposition under AW activation was investigated for DS experiments. **Figure 2 a)** shows a series of photographs of samples prepared for the indicated times under AW activation and, for comparison, simultaneously without activation. In the two cases, the samples depicted a homogenous aspect without any hint of patterning in the color distribution. A close inspection of the coloring of the two sets of samples reveals that $D_{ref}$ samples are darker than those prepared under AW activation for a given period of time. This coloring is typical of deposited silver NPs and is attributed to their plasmon resonance absorption.[28–30,47,48] The spectra in **Figure 2b)** and **2c)** correspond to depositions for 15, 30 and 40 min and clearly reflect a progressive variation in the plasmon spectra of the samples. The absorption bands are better defined, i.e. narrower peaks, and have minima at shorter wavelengths in the samples prepared under AW activation (minima of the plasmon band at 483±5, 476±5, and 540±5 nm for samples D15-D40, vs 502±5 and 580±5 nm for samples $D_{ref}15$, $D_{ref}30$, and not well-defined plasmon in samples $D_{ref}40$). From the common knowledge about the plasmon properties of plasmonic nanoparticles, this behavior suggests that nanoparticles are smaller, more homogeneous in size, and present a lesser coalescence in the AW-activated samples than in the reference ones.

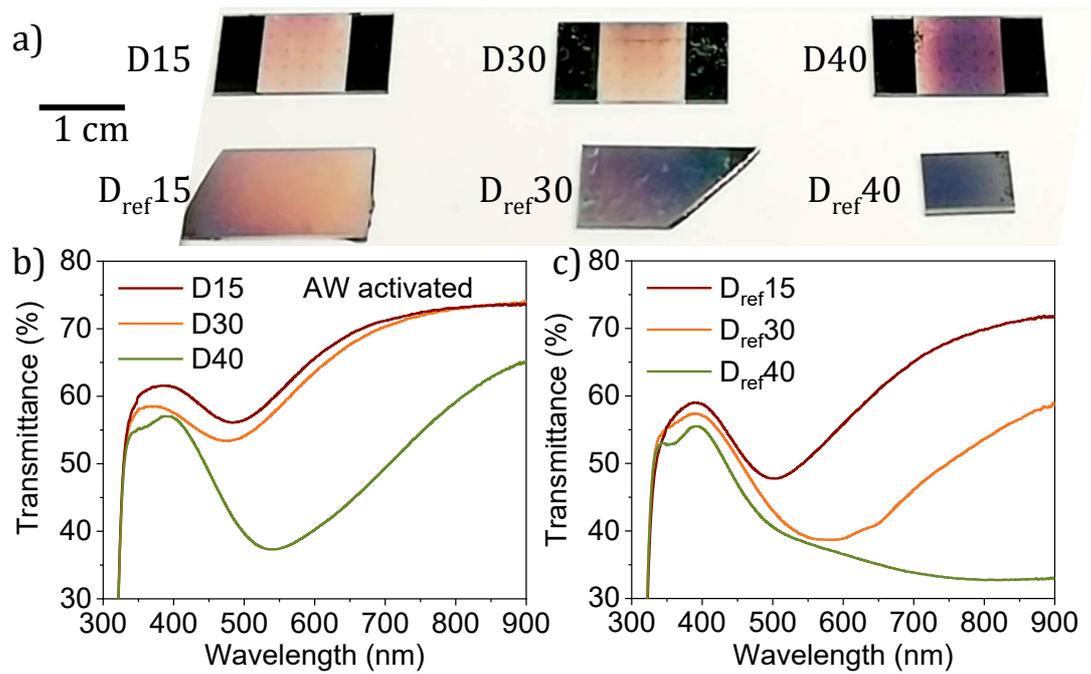

*Figure 2.- AW activated deposition of silver nanoparticles by diode DC sputtering. a) Photographs of AW-activated samples and comparison with reference samples prepared simultaneously. b) UV-vis absorption spectra taken for samples D15-D40. c) Idem for $D_{ref}15$-$D_{ref}$-40 samples.*

SEM analysis confirmed this assessment of the morphological characteristics of the NPs. **Figure 3** shows a series of normal-view SEM micrographs and the contours obtained by applying the *ImageJ* software for samples deposited for 15, 30 and 40 minutes.

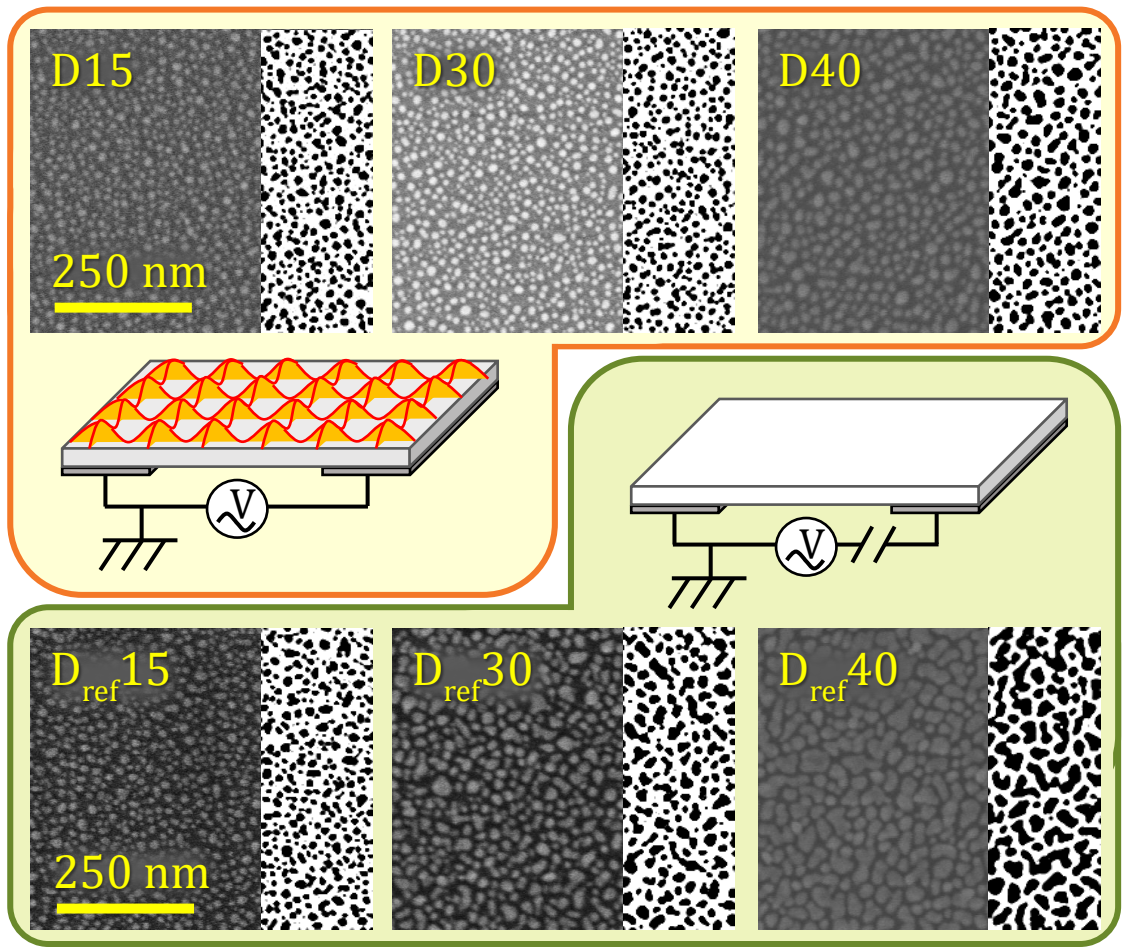

*Figure 3.- SEM characterization and comparison of reference and AW-activated deposited Ag nanoparticles by DS. SEM micrographs (left) and J-image derived contours (right) of samples D15-D40 (top) and $D_{ref}15$-$D_{ref}40$ (bottom).*

At first sight, it appears that NPs are bigger in samples $D_{ref}$ than in D and that an incipient aggregation has occurred in samples $D_{ref}30$ and $D_{ref}40$. This assessment was fully confirmed by analyzing the particle size distribution and circularity with *ImageJ* software. **Figure 4** shows the evolution of these two magnitudes for samples D15-D40 and $D_{ref}15$-$D_{ref}40$. The mean particle size and circularity values deduced from this analysis are summarized in **Table 1**. An estimate of the percentage of the substrate surface covered by silver and the number of particles in an area of $5 \cdot 10^5$ $nm^2$ (area of a micrograph) are also included in this table. Both Figure 4 and Table 1 prove that silver NPs are smaller and more circular in the AW-activated samples. Data in Table 1 also show that the number of NPs per unit area is higher in the AW-activated samples (up to 2.6 times higher for 40 min deposited samples). Interestingly, the percentage of substrate area covered by silver is rather similar in all cases except for sample $D_{ref}40$ where coverage amounts to almost half of the substrate. This coincides with a high degreee of agglomeration of NPs, which are the largest and the less circular of the whole series.

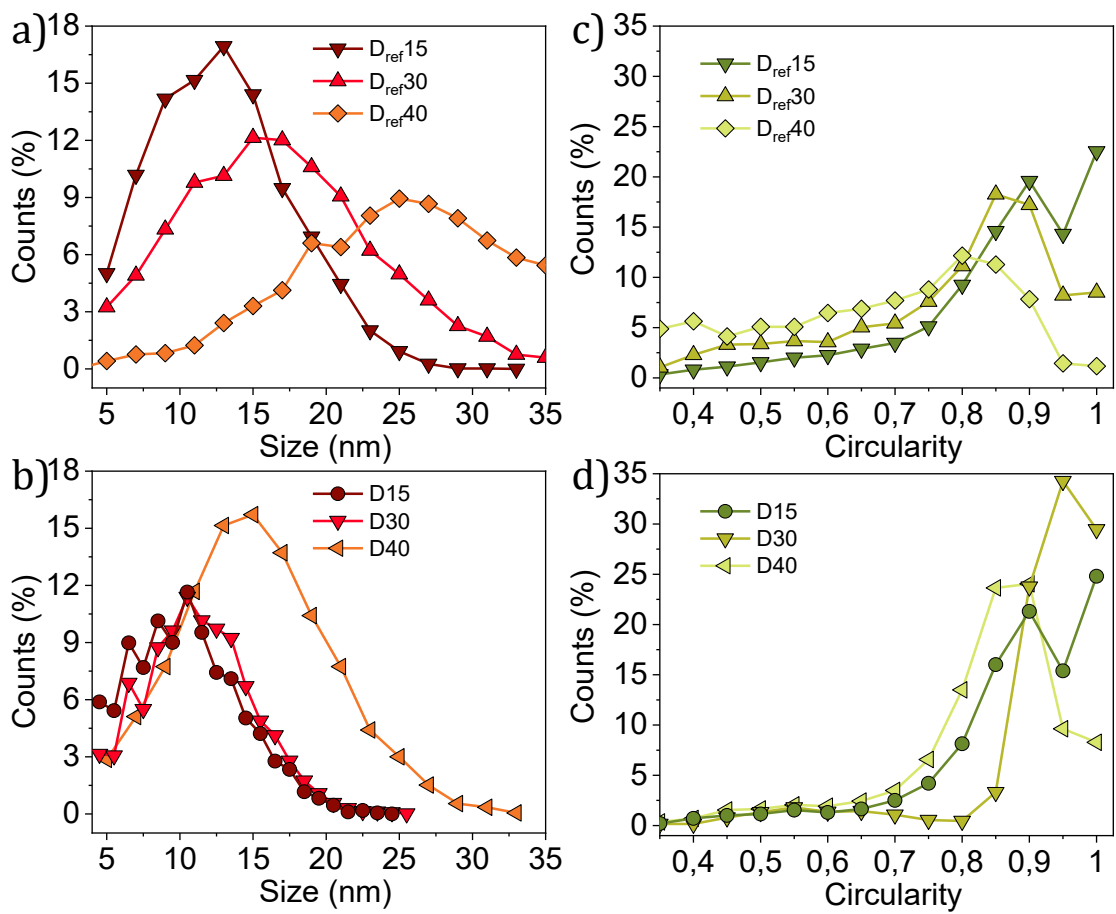

*Figure 4.- Statistical analysis of DS nanoparticles morphology. a) Particle size distribution plots for samples $D_{ref}15$, $D_{ref}30$, and $D_{ref}40$(left) D15, D30, and D40(right), as indicated. b) Idem circularity.*

*Table 1.- Mean particle size and circularity for samples D and MS*

|  | $D15/D_{ref}15$ | $D30/D_{ref}30$ | $D40/D_{ref}40$ | $MS1_b/MS1_w$* | $MS_{ref}1$ | $MS2_b/MS2_w$* | $MS_{ref}2$ |
|---|---|---|---|---|---|---|---|
| *Mean size (nm)* | 10.4/12.9 | 11.4/16.9 | 15.2/29.3 | 9.5/14.3 | 21.2 | 11.6/28.5 | 42.5 |
| *Circularity* | 0.90/0.88 | 0.94/0.80 | 0.85/0.65 | 0.85/0.72 | 0.77 | 0.88/0.72 | 0.83 |
| *Covered area (%)* | 23.0/24.6 | 24.7/34.4 | 32.5/49.9 | 19.5/28.8 | 46.4 | 15.8/52.3 | 52.4 |
| *Number of particles in $5 \cdot 10^5$ nm$^2$* | 5476/5303 | 5091/3056 | 3697/1454 | 1097/715 | 446 | 879/265 | 290 |
| *Particle Ratio AW/Ref* | 1.03 | 1.66 | 2.53 | 2.46/1.60 | 1 | 3.03/0.91 | 1 |

*The two values provided in these columns correspond to the bluish ($MS_b$) and whitish ($MS_w$) domains developed in MS samples.*

### Silver NP formation by MS deposition under AW activation.

Since the deposition rate by MS is much higher than by DS deposition, much shorter deposition times, on the order of 1-2 min were required to achieve a similar coverage in both cases. For samples MS1 and MS$_{ref}$1, **Figure 5** shows some basic characterization results including their photographs (**Figure 5a**), a specific image treatment of the color distribution (**Figure 5b**), and the UV-vis absorption spectra of these two samples taken with a standard spectrometer with a beam approximately 5mm in diameter (**Figure 5c**). This figure highlights that unlike the homogeneous character of sample MS$_{ref}$1, sample MS1 presents on a blue background a patterned distribution of whitish spots aligned perpendicularly to the pattern of standing AW and separated by a distance of ca. 1 mm. This patterned distribution of standing AWs has been characterized by doppler vibrometry and simulated by COMSOL in previous works. [17,18]

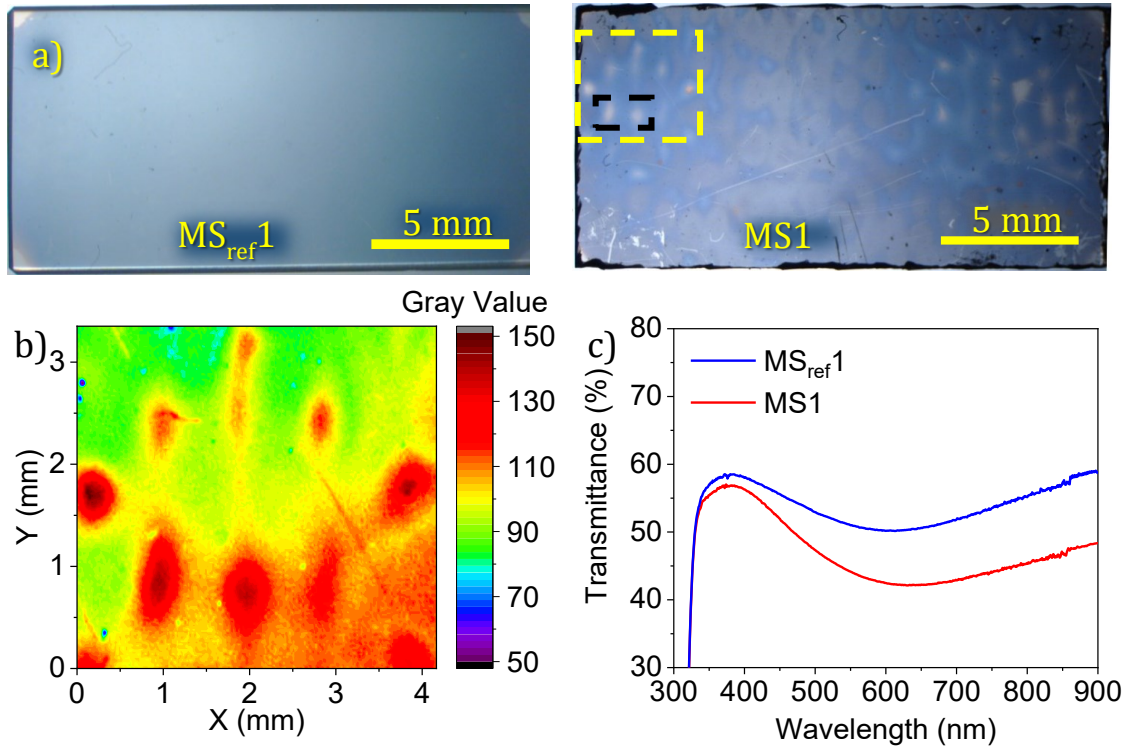

*Figure 5.- a) Photographs of samples MS1 and MS$_{ref}$1, as indicated. b) Color map of the substrate area framed within the yellow dashed square of the MS1 photo in a). The color scale represents the grey intensity value after converting the image to a grey scale, the yellow-green color representing the greyest and the dark-red the whitest. c) UV-vis absorption spectra of these two samples.*

The color map of the grey scale distribution (Figure 5b) confirms that the spots are laterally separated by a distance of approximately 1 mm. This pattern is very similar to the standing AW pattern generated onto the LiNbO$_3$ substrate as reported in ref.[17,18]. The homogenous bluish coloration of sample M$_{ref}$1 is due to a relatively broad plasmon absorption band at 608 nm (Figure 5c), i.e., close to the minimum of the absorption band of sample D$_{ref}$40. Meanwhile, the plasmon absorption spectrum of sample MS1 appears at 633 nm and must correspond to an average of the contributions of the bluish and whitish zones examined with the spectrometer. As a working hypothesis, we assume that this patterned distribution of coloration reflects a heterogeneous distribution of silver NPs on the substrate.

A detailed characterization of the particle size distribution in these samples was obtained by SEM. **Figure 6,** shows selected SEM micrographs of samples MS$_{ref}$1 and MS1, in this case reporting a different distribution of silver NPs in the bluish and whitish zones developed in this sample. In this figure, the SEM images and the contour plots obtained by the ImageJ software prove that NPs are bigger in sample $MS_{ref}1$ where a certain agglomeration of NPs has already taken place. In sample MS1, it is also apparent that the NP size is smaller in the whitish spots than in the bluish zones. A more complete set of SEM results for sample MS1 is reported in the supporting information section, **Figure S5,** where images taken at 100 µm intervals along a line connecting the whitish and the bluish domains are reported. A similar evolution of particle sizes and shapes between the whitish and bluish domains are found for the all regions of the sample (see supporting information **S6** for the analysis of other zones, as indicated).

Complementary evidence about the distribution of silver NPs in the whitish and bluish domains of sample MS1 refers to the percentage of substrate surface covered by silver in each type of domain (this magnitude can be taken as a semiquantitative estimate of the amount of silver). An evaluation of this parameter obtained by the particle contour maps in Figure 6 renders percentages of *19.47* and *28.75* for sample MS1 (whitish, $MS1_w$, and bluish, $MS1_b$, zones) and *46.37* for sample $MS_{ref}1$. The distinct values found for sample MS1 suggest that there is less silver on the whitish domains of this sample. This assessment was confirmed by a linear EDX analysis of the evolution of the silver content, giving an approximate ratio of *4.0/7.3* ($MS1_w/MS1_b$) for the silver EDX signals taken in these two zones.

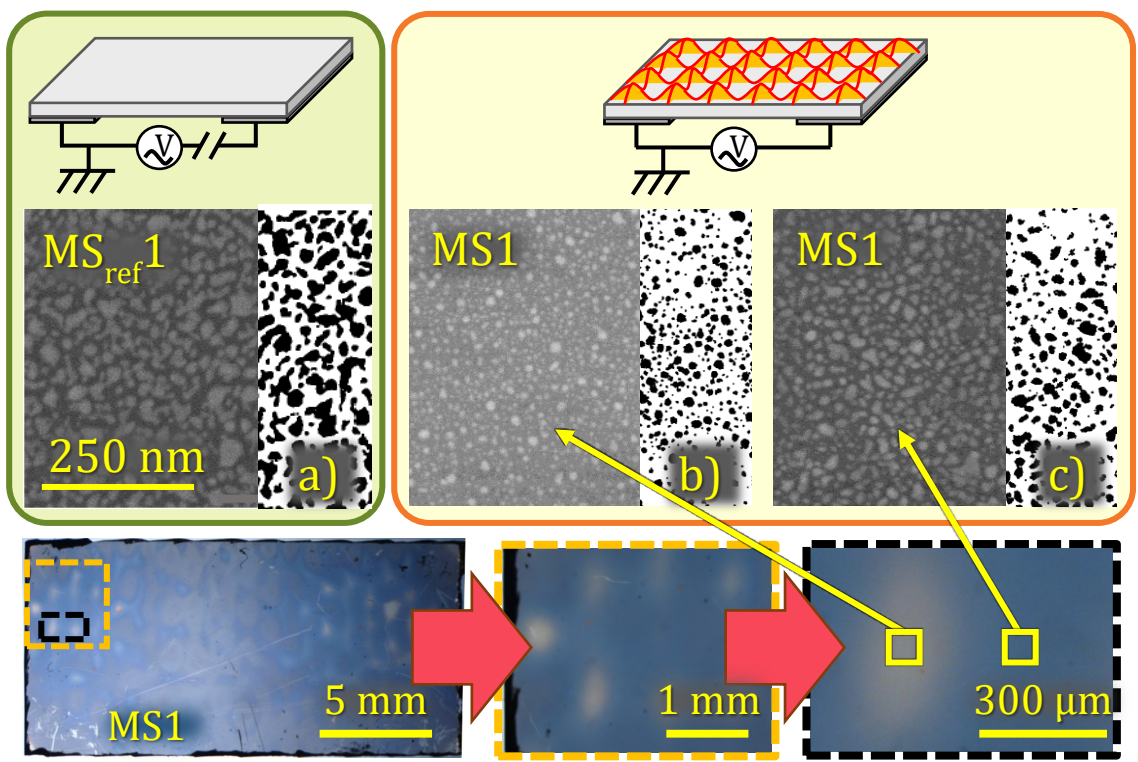

*Figure 6.- SEM micrographs (left) and contour maps (right) obtained with the imaging software ImageJ for samples MS$_{ref}$1(a) and MS1(b, c). The two images/contour maps of sample MS1 corresponding to the center of a whitish zone and the adjacent bluish zone have been taken in the zone defined by the black dashed square of the MS1 photo in Figure 5 a). The same photo is reproduced in the bottom part of this figure where it is presented in successive enlargements from left to right.*

The particle size distribution and circularity of the silver NPs determined from the analysis of the images in **Figure 6** are reported in **Figure 7a**. The average value of particle sizes and circularities included in Table 1 clearly confirm that the size of NPs is smaller in the MS1 sample than in the MS$_{ref}$1. Moreover, the circularity analysis with a clear maximum at *0.85/0.80* (Figure 7) and an average value of *0.85/0.72* (Table 1) for the NPs in the whitish/bluish domains in sample MS1 further confirms that NPs are different than in sample M$_{ref}$1, where a broad distribution of circularity with values down to *0.70* (Figure 7b) and an average value of *0.77* are found. On the other hand, the plots in Figure 7b showing the variation in average particle size and circularity along a line connecting a whitish and a bluish zone in MS1 depict a size variation from ca. *10* nm to *15* nm and a circularity change from *0.90* to *0.70*. A similar evolution is found for the NPs in adjacent domains in other regions of the substrate (see for example the analysis reported in the supporting information **Figure S5 and S6**).

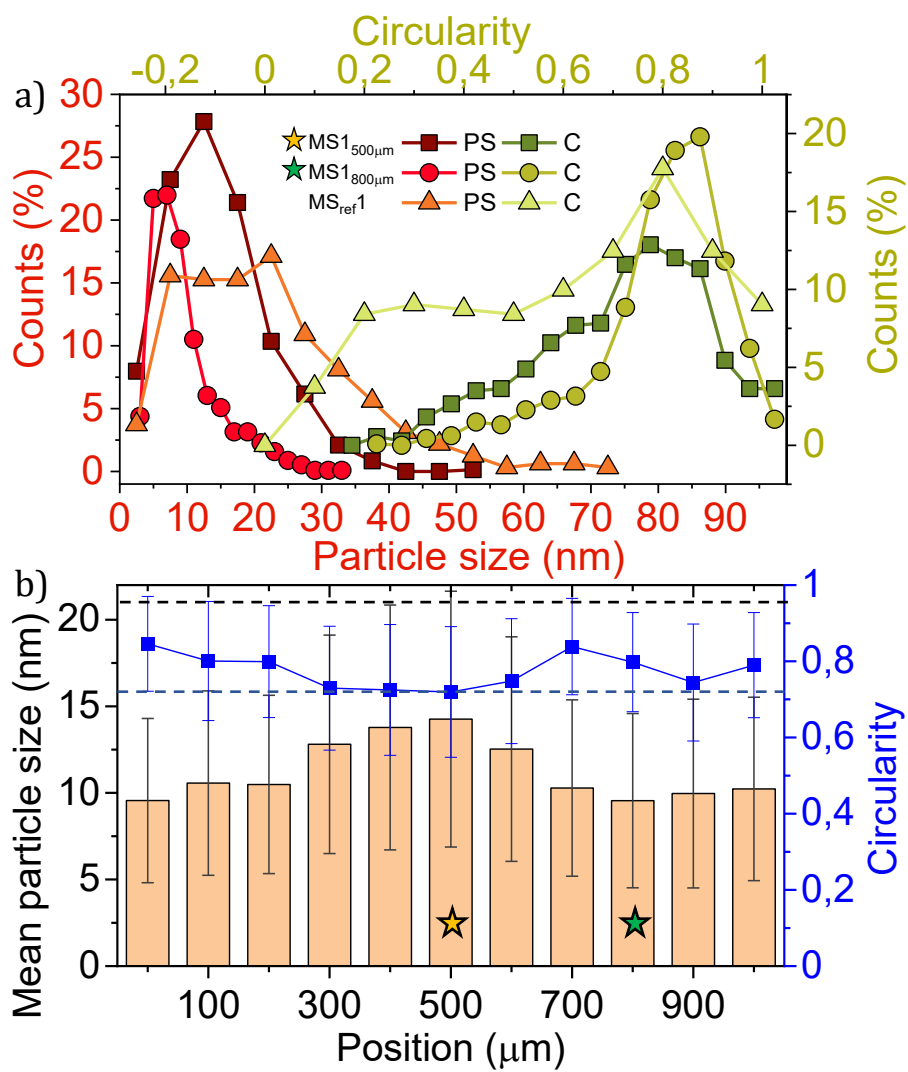

*Figure 7.- a) Particle size and circularity distribution of the particles of a bluish and a whitish region of sample MS1 (positions 500 and 800nm, and yellow and green stars in panel b), respectively). The values for the MS$_{ref}$1 are also included for comparison.*

*b) Average Particle size distribution and circularity of the silver NPs along a path connecting the two whitish regions contained in the black dashed square of the MS1 photo in Figure 5 a). The black and blue dashed lines represent the mean particle size and the circularity of the sample MS$_{ref}$1, respectively.*

An analysis similar to that performed for sample MS1 was also done for a MS sample deposited for 2 min (MS2 and MS$_{ref}$2). The particle size distribution and circularity results, presented in the supporting information section (**Figure S7**) and the average values of these parameters included in **Table 1** show similar tendencies than for samples MS1-MS$_{ref}$1, i.e. larger and less rounded particles in sample MS$_{ref}$2 than in MS2, and in this latter case, smaller and more rounded particles in the whitish domains compared to the bluish zones.

**Discussion**

In previous work we have shown that the lateral field electroacoustic activation of a piezoelectric LiNbO$_3$ substrate with an AC signal at a frequency around 3.4-3.5 MHz gives rise to the generation of plate standing bulk AWs with a predominant shear character[17,18] and a wavelength on the order of 1 mm. The results above have evidenced that this activation of the substrate deeply affects the initial stages of the plasma deposition of silver and that the accumulation of metal up to the full coverage of the substrate hampers the activation. In fact, according to Figure 1, S$_{11}$ intensity and frequency remain rather constant during the first minutes of silver deposition, when no silver agglomeration neither a continuous film formation have yet taken place. A shift in resonance frequency and the broadening and eventual vanishing of the S$_{11}$ spectrum occur for deposition times longer than 7 min. This behavior is consistent with the agglomeration of silver NPs into a continuous silver layer on top of the piezoelectric substrate, a situation that precludes the formation and propagation of AWs. This effect has been suggested as a monitoring procedure of a metallic thin film formation in other types of TSM devices .[46]

The results in Figures 2-7 have revealed that for short deposition times silver forms separated NPs, whose shape and distribution differ from those found in reference samples. This proves that the AWs affect the initial stages of silver deposition.

A general result of the AW activation during deposition is that NP size was smaller, circularity higher, and their agglomeration degree smaller than in reference samples. By MS deposition, a characteristic submillimeter patterning of the NP distribution has also been found. These differences had a direct correlation with the plasmon resonance spectra as shown in Figures 2 and 5.

Nucleation and growth of silver NPs by sputtering have been widely studied either experimentally[49–51] or by theoretical simulation.[52,53] These studies provide a clear understanding of the individual steps involved in this process, namely ad-atom diffusion, nuclei formation, particle growth, and particle agglomeration.[25–27] These stages, except for the ad-atom diffusion that cannot be observed in our experiment, can be clearly recognized in the series of SEM micrographs in Figures 3 and 6 for the $D_{ref}$ and $MS_{ref}$ samples. Significantly, the AW activation of sputtered NP growth seems to prevent or delay the agglomeration stages, which cannot be easily identified for samples D and MS in Figures 3 and 6. There may be several factors responsible for this difference. The first hypothesis is that the mechanical oscillations induced in the $LiNbO_3$ substrate due to the inverse piezoelectric effect provide momentum and kinetic energy to the ad-atoms, nuclei, and NPs on the surface, this leading to an increase in mobility. The considerable shear component (i.e. around one order of magnitude larger than the normal component)[9] of the generated standing AWs might be particularly effective in this regard. For the DS experiment and long deposition times, it can neither be discarded that acousto-thermal effects[54] might contribute to enhance the diffusion of ad-atoms and nuclei. However, the experimental trends appear contradictory with an enhancement of mobility, since in all AW-activated samples the NPs were more numerous, smaller in size, and more circular than in the reference ones. This suggests either that the mechanical oscillations of AWs are ineffective in mobilizing silver or, most likely, that additional factors counteract the mobilization or diffusion of species and/or induce the disruption of the nuclei formed during the initial stages of the deposition process.

A rough estimation of the percentage of substrate area covered by silver renders values that were approximately 35% smaller in the AW-activated samples (cf. Table 1). This feature, together with the relatively less amount of silver determined by EDX analysis in the AW-activated samples, indicates that the effective sticking coefficient of silver is smaller. This is to say that some ad-atoms might have been reemitted from the surface under the combined effect of plasma and AW excitation.

Since AW activation was effective in altering NP formation and distribution only if applied during deposition (i.e., when ad-atoms become adsorb and diffuse onto the surface), the aforementioned changes with respect to reference samples should consider the interaction of plasma with the activated substrate as the main activation phenomenon. In our previous work on the AW-activated MS deposition of dielectric layers, we proved that the main factor accounting for the 2D patterning of the films was a focused bombardment with accelerated Ar$^+$ plasma ions, an effect that was due to the polarization field generated at the piezoelectric substrate during excitation. [17,18] We claim a similar mechanism to account for the distinct NP growth found in this work for the AW-activated samples. The simulation model of this process in this previous work showed that the energy of impinging ions onto the AW-activated piezoelectric substrate may reach values in the order of 10 eV or higher for MS deposition conditions. This energy range is high enough to provoke the reemission of weakly surface-bonded silver ad-atoms, the formation of surface defects on the piezoelectric plate acting as nucleation centers, and the prevention of coalescent phenomena among NPs.[55–58] A schematic of this focused ion bombardment process during MS deposition is shown in **Figure 8a)**. Therefore, the effective focusing of relatively high energy ions in samples MS1/MS2 must be blamed for the 2D patterning observed in the distribution of NPs. The effects are more relevant in the whitish zone where ion impingement is maximum. The effect of this local exaltation of ion bombardment induces a further decrease in the size of NPs and in the amount of silver, both phenomena resulting from a localized sputtering/reemission of ad-atoms.

The main difference between samples D and MS is the absence of any 2D patterning in the former case, although they still present similar differential tendencies when comparing the features of their NPs with those in samples $D_{ref}$. In a recent work on the plasma characteristics of a cylindrical diode sputtering discharge[35] it is reported that while the electron density, on the order of $10^{15}$ m$^{-3}$, does not significantly differ from values typical of a MS discharge, the electron energy sharply drops at distances higher than 2-3 cm from the cathode. Moreover, simulation analysis of this kind of plasma renders that its characteristic Debye length, and therefore the thickness of the sheath formed on the substrate, is around ten times larger than in a typical MS configuration. This fact, together with the higher gas pressure (and therefore shorter mean free path) when operating the DS (1.7· 10$^{-1}$ mbar) with respect to the MS (5· 10$^{-3}$ mbar) discharges, are two factors that may account for the lack of patterning phenomena in samples D. Under these conditions, the much thicker sheath formed over the substrate during DS has a collisional character (a rough estimate renders that the mean free path of species is $\lambda_{mfp} \approx 0.3$ mm against a Debye length of $\lambda_D \approx 1$ mm away from the cathode).[35,36] This means that ions pervasively collide with the gas species in their trajectories through the sheath up to the substrate where they arrive along randomized trajectories and comparatively less energy. This agrees with that no effective focusing occurred in samples D, in spite that the activation voltage of the AW was 112 V (for 10 V by MS deposition) and that the polarization field at the surface of the piezoelectric plate should be larger than during deposition of MS samples. These differences between MS and D processes are schematized in Figure 8, showcasing that no 2D patterning appears in the latter case.

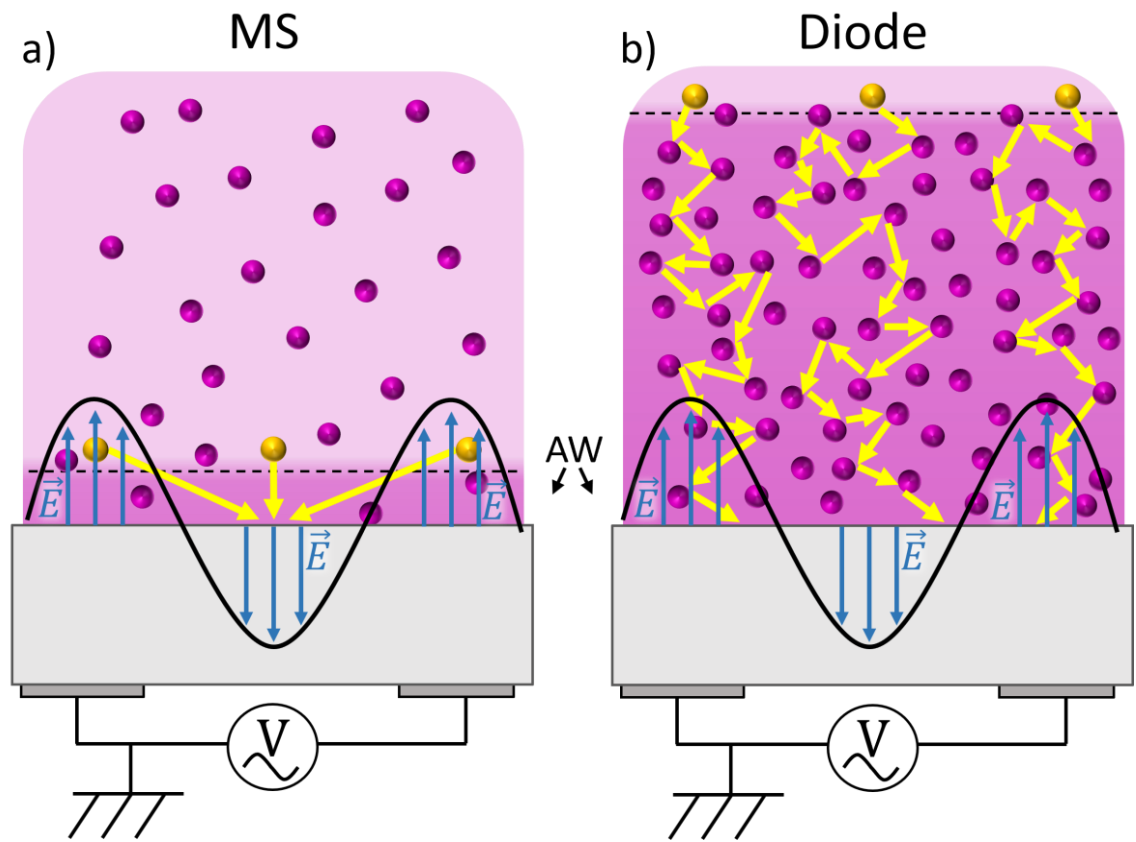

*Figure 8.- Schematic of the sheaths formed around the substrates during deposition in MS a) and D b). The yellow particles and the arrows depict hypothetical trajectories of ions that impinge on the substrate. The boundary limits of the sheaths are indicated by black dashed lines in both cases. The sinusoidal wave in the two cases represents the local polarization of the substrate due to its AW activation.*

In summary, the findings reported in this work about the formation of silver NPs by sputtering deposition on AW-activated substrates can be accounted for by the occurrence of Ar+ ion bombardment phenomena, as previously modeled for dielectric thin film deposition.[18] The characteristics of the plasma (electron energy and pressure) may play a large role in this regard, since focusing effects can be neglected for certain plasma conditions.

## Conclusions

The previous results and discussion have shown that following the evolution of the AW resonances of piezoelectric substrates during the plasma deposition of metals provides a monitoring procedure to assess the formation and agglomeration of particles. It has also been found that NPs formed by plasma deposition techniques under AW activation differ in size, shape, and agglomeration state from those prepared without AW excitation. This provides an effective way to tailor particle sizes and shapes, provided that the deposition/activation processes are well controlled. The interaction of the electrical polarization field generated at the substrate with the plasma ignited during MS deposition gives rise to a 2D pattern where nucleation and growth of sputtered NPs occur differently due to the interaction with focused and accelerated Ar$^+$ plasma ions. We prove that these focused ion impingement phenomena do not occur during diode sputtering, where plasma characteristics and gas pressure preclude achieving the conditions required for the focusing phenomena to occur. The observable differences in NP size and circularity with respect to reference samples must stem from a combination of excitation phenomena including a residual ion bombardment, the mechanical activation of ad-atoms and/or acousto-thermal effects.

In general, the modulation of the sputtering deposition by AWs can be used as an efficient mechanism to retard the agglomeration stage of NPs, a typical step by the deposition of silver thin films. This permits the fabrication of NPs with a higher circularity and smaller size than when the deposition is carried out without AW activation. A 2D patterning and the formation of domains, where NPs depict different sizes and circularities, is another output result of this activation by AWs.

**Acknowledgments**

The authors acknowledge the projects PID2020-112620GB-I00, and PID2020-114270RA-I00 funded by MCIN/AEI/10.13039/501100011033, the project TED2021-130124A-I00 funded by AEI/10.13039/501100011033/Unión Europea Next Generation EU/PRTR and the EU H2020 program under grant agreement 899352 (FETOPEN-01-2018-2019-2020 – SOUNDofICE). M. Oliva-Ramírez acknowledges financial support from Grant IJC2020-045087-I funded by: MCIN/ AEI /10.13039/501100011033 and the European Union NextGeneration EU /PRTR. Guillermo Regodón acknowledges funding from the European Commission-NextGenerationEU through the "Plan de Recuperación, Transformación y Resiliencia" from the Spanish Government.

**References**

1. L. Y. Yeo and J. R. Friend, Annual Review of Fluid Mechanics **46**, 379 (2014).

2. S. P. Zhang, J. Lata, C. Chen, J. Mai, F. Guo, Z. Tian, L. Ren, Z. Mao, P.-H. Huang, P. Li, S. Yang, and T. J. Huang, Nat Commun **9**, 2928 (2018).

3. H. Sazan, S. Piperno, M. Layani, S. Magdassi, and H. Shpaisman, Journal of Colloid and Interface Science **536**, 701 (2019).

4. D. J. Collins, B. Morahan, J. Garcia-Bustos, C. Doerig, M. Plebanski, and A. Neild, Nat Commun **6**, 8686 (2015).

5. A. Salari, S. Appak-Baskoy, M. Ezzo, B. Hinz, M. C. Kolios, and S. S. H. Tsai, Small **16**, 2004323 (2020).

6. N. Zhang, A. Horesh, and J. Friend, Advanced Science **8**, 2100408 (2021).

7. L. Shu, S. Ke, L. Fei, W. Huang, Z. Wang, J. Gong, X. Jiang, L. Wang, F. Li, S. Lei, Z. Rao, Y. Zhou, R.-K. Zheng, X. Yao, Y. Wang, M. Stengel, and G. Catalan, Nat. Mater. **19**, 605 (2020).

8. G. Ji and J. Huber, Applied Materials Today **26**, 101260 (2022).

9. J. del Moral, L. Montes, V. J. Rico-Gavira, C. López-Santos, S. Jacob, M. Oliva-Ramirez, J. Gil-Rostra, A. Fakhfouri, S. Pandey, M. Gonzalez del Val, J. Mora, P. García-Gallego, P. F. Ibáñez-Ibáñez, M. A. Rodríguez-Valverde, A. Winkler, A. Borrás, and A. R. González-Elipe, Advanced Functional Materials **33**, 2209421 (2023).

10. S. Jacob, S. Pandey, J. D. Moral, A. Karimzadeh, J. Gil-Rostra, A. R. González-Elipe, A. Borrás, and A. Winkler, Advanced Materials Technologies **8**, 2300263 (2023).

11. D. Yang, R. Tao, X. Hou, H. Torun, G. McHale, J. Martin, and Y. Fu, Advanced Materials Interfaces **8**, 2001776 (2021).

12. P. Delsing, A. N. Cleland, M. J. A. Schuetz, J. Knörzer, G. Giedke, J. I. Cirac, K. Srinivasan, M. Wu, K. C. Balram, C. Bäuerle, T. Meunier, C. J. B. Ford, P. V. Santos, E. Cerda-Méndez, H. Wang, H. J. Krenner, E. D. S. Nysten, M. Weiß, G. R. Nash, L.

Thevenard, C. Gourdon, P. Rovillain, M. Marangolo, J.-Y. Duquesne, G. Fischerauer, W. Ruile, A. Reiner, B. Paschke, D. Denysenko, D. Volkmer, A. Wixforth, H. Bruus, M. Wiklund, J. Reboud, J. M. Cooper, Y. Fu, M. S. Brugger, F. Rehfeldt, and C. Westerhausen, J. Phys. D: Appl. Phys. **52**, 353001 (2019).

13. W. Shen, M. Wang, X. Sun, G. Liu, Z. Li, and S. Liu, Microchemical Journal **180**, 107576 (2022).

14. Y. Zhang, F. Yang, Z. Sun, Y.-T. Li, and G.-J. Zhang, Analyst **142**, 3468 (2017).

15. M. Takahashi, H. Shoji, and M. Tsunoda, Journal of Magnetism and Magnetic Materials **134**, 403 (1994).

16. A. Romanyuk, V. Spassov, and V. Melnik, Journal of Applied Physics **99**, 034314 (2006).

17. A. García-Valenzuela, A. Fakhfouri, M. Oliva-Ramírez, V. Rico-Gavira, T. C. Rojas, R. Alvarez, S. B. Menzel, A. Palmero, A. Winkler, and A. R. González-Elipe, Mater. Horiz. **8**, 515 (2021).

18. V. Rico, G. F. Regodón, A. Garcia-Valenzuela, A. M. Alcaide, M. Oliva-Ramirez, T. C. Rojas, R. Alvarez, F. J. Palomares, A. Palmero, and A. R. Gonzalez-Elipe, Acta Materialia **255**, 119058 (2023).

19. J. L. Rose, *Ultrasonic Guided Waves in Solid Media* (Cambridge University Press, Cambridge, 2014).

20. A. Demaude, K. Baert, D. Petitjean, J. Zveny, E. Goormaghtigh, T. Hauffman, M. J. Gordon, and F. Reniers, Advanced Science **9**, 2200237 (2022).

21. N. Kaiser, Appl. Opt., AO **41**, 3053 (2002).

22. Z. Zhang and M. G. Lagally, Science **276**, 377 (1997).

23. C. Ratsch and J. A. Venables, Journal of Vacuum Science & Technology A **21**, S96 (2003).

24. G. Abadias, E. Chason, J. Keckes, M. Sebastiani, G. B. Thompson, E. Barthel, G. L. Doll, C. E. Murray, C. H. Stoessel, and L. Martinu, Journal of Vacuum Science & Technology A **36**, 020801 (2018).

25. V. Elofsson, B. Lü, D. Magnfält, E. P. Münger, and K. Sarakinos, Journal of Applied Physics **116**, 044302 (2014).

26. V. Gervilla, G. A. Almyras, F. Thunström, J. E. Greene, and K. Sarakinos, Applied Surface Science **488**, 383 (2019).

27. V. Gervilla, G. A. Almyras, B. Lü, and K. Sarakinos, Sci Rep **10**, 2031 (2020).

28. A. N. Filippin, A. Borras, V. J. Rico, F. Frutos, and A. R. González-Elipe, Nanotechnology **24**, 045301 (2013).

29. J. Okumu, C. Dahmen, A. N. Sprafke, M. Luysberg, G. von Plessen, and M. Wuttig, Journal of Applied Physics **97**, 094305 (2005).

30. T. W. H. Oates and A. Mücklich, Nanotechnology **16**, 2606 (2005).

31. J. Wang, K. M. Koo, Y. Wang, and M. Trau, Advanced Science **6**, 1900730 (2019).

32. K. Hurtuková, K. Fajstavrová, S. Rimpelová, B. Vokatá, D. Fajstavr, N. S. Kasálková, J. Siegel, V. Švorčík, and P. Slepička, Materials **14**, 4051 (2021).

33. A. Costas, N. Preda, I. Zgura, A. Kuncser, N. Apostol, C. Curutiu, and I. Enculescu, Sci Rep **13**, 10698 (2023).

34. B. Pucelik, A. Sułek, M. Borkowski, A. Barzowska, M. Kobielusz, and J. M. Dąbrowski, ACS Appl. Mater. Interfaces **14**, 14981 (2022).

35. T. Richard, I. Furno, A. Sublet, and M. Taborelli, Plasma Sources Sci. Technol. **29**, 095003 (2020).

36. J. T. Gudmundsson and D. Lundin, in *High Power Impulse Magnetron Sputtering*, edited by D. Lundin, T. Minea, and J. T. Gudmundsson (Elsevier, 2020), pp. 1–48.

37. I. Petrov, V. Orlinov, I. Ivanov, and J. Kourtev, Contributions to Plasma Physics **28**, 157 (1988).

38. A. Barranco, A. Borras, A. R. Gonzalez-Elipe, and A. Palmero, Progress in Materials Science **76**, 59 (2016).

39. B. Bouaouina, C. Mastail, A. Besnard, R. Mareus, F. Nita, A. Michel, and G. Abadias, Materials & Design **160**, 338 (2018).

40. M. R. FitzGerald, A Lateral Field Excited Thin Film Bulk Acoustic Wave Sensor, Maine, 2013.

41. Y. Yang, R. Lu, T. Manzaneque, and S. Gong, in *2018 IEEE International Frequency Control Symposium (IFCS)* (2018), pp. 1–5.

42. J. S. R. Hartz, N. W. Emanetoglu, C. Howell, and J. F. Vetelino, Biointerphases **15**, 030801 (2020).

43. V. Plessky, S. Yandrapalli, P. j. Turner, L. g. Villanueva, J. Koskela, and R. b. Hammond, Electronics Letters **55**, 98 (2019).

44. C. Zhao, W. Geng, X. Qiao, F. Xue, J. He, G. Xue, Y. Liu, H. Wei, K. Bi, Y. Li, M. Xun, and X. Chou, Sensors and Actuators A: Physical **333**, 113230 (2022).

45. C. A. Schneider, W. S. Rasband, and K. W. Eliceiri, Nature Methods **9**, 671 (2012).

46. N. Yoshimura, N. Nakamura, H. Ogi, and M. Hirao, in *Proceedings of Symposium on Ultrasonic Electronics* (Kyoto (Japan), 2013), pp. 195–196.

47. J. R. Sanchez-Valencia, J. Toudert, A. Borras, A. Barranco, R. Lahoz, G. F. de la Fuente, F. Frutos, and A. R. Gonzalez-Elipe, Adv. Mater. **23**, 848 (2011).

48. V. Amendola, R. Pilot, M. Frasconi, O. M. Maragò, and M. A. Iatì, J. Phys.: Condens. Matter **29**, 203002 (2017).

49. A. Shelemin, P. Pleskunov, J. Kousal, J. Drewes, J. Hanuš, S. Ali-Ogly, D. Nikitin, P. Solař, J. Kratochvíl, M. Vaidulych, M. Schwartzkopf, O. Kylián, O. Polonskyi, T. Strunskus, F. Faupel, S. V. Roth, H. Biederman, and A. Choukourov, Particle & Particle Systems Characterization **37**, 1900436 (2020).

50. J. Zhao, E. Baibuz, J. Vernieres, P. Grammatikopoulos, V. Jansson, M. Nagel, S. Steinhauer, M. Sowwan, A. Kuronen, K. Nordlund, and F. Djurabekova, ACS Nano **10**, 4684 (2016).

51. P. Asanithi, S. Chaiyakun, and P. Limsuwan, Journal of Nanomaterials **2012**, e963609 (2012).

52. P. Grammatikopoulos, Current Opinion in Chemical Engineering **23**, 164 (2019).

53. J. Vernieres, S. Steinhauer, J. Zhao, A. Chapelle, P. Menini, N. Dufour, R. E. Diaz, K. Nordlund, F. Djurabekova, P. Grammatikopoulos, and M. Sowwan, Advanced Functional Materials **27**, 1605328 (2017).

54. J. Kondoh, N. Shimizu, Y. Matsui, and S. Shiokawa, Ultrasonics, Ferroelectrics and Frequency Control, IEEE Transactions On **52**, 1881 (2005).

55. J. Wang, R. Shu, J. Chai, S. G. Rao, A. le Febvrier, H. Wu, Y. Zhu, C. Yao, L. Luo, W. Li, P. Gao, and P. Eklund, Materials & Design **219**, 110749 (2022).

56. A. R. González-Elipe, F. Yubero, and J. M. Sanz, *Low Energy Ion Assisted Film Growth* (PUBLISHED BY IMPERIAL COLLEGE PRESS AND DISTRIBUTED BY WORLD SCIENTIFIC PUBLISHING CO., 2003).

57. I. Utke, P. Hoffmann, and J. Melngailis, Journal of Vacuum Science & Technology B: Microelectronics and Nanometer Structures Processing, Measurement, and Phenomena **26**, 1197 (2008).

58. H. Ma, Y. Zou, A. S. Sologubenko, and R. Spolenak, Acta Materialia **98**, 17 (2015).